\documentclass[fleqn,usenatbib]{mnras}

\usepackage{newtxtext,newtxmath}
\usepackage[T1]{fontenc}
\usepackage{ae,aecompl}

\usepackage{graphicx}	% Including figure files
\usepackage{amsmath}	% Advanced maths commands
\usepackage{amssymb}	% Extra maths symbols
\usepackage{multirow}

\title[Scientific performance analysis of SYZ vs. RC telescopes]{Scientific performance analysis of the SYZ telescope design vs. the RC telescope design }

\author[Donglin Ma \& Zheng Cai]{
Donglin Ma,$^{1}$
and Zheng Cai$^{2,3,}$\thanks{E-mail: zcai@ucolick.org (Corresponding author)}
\\
$^{1}$School of Optical and Electronic Information, Huazhong University of Science and Technology, Wuhan, 430074, China\\
$^{2}$UCO/Lick Observatory, University of California at Santa Cruz, 1156 High Street, Santa Cruz, CA 95064, USA\\
$^{3}$Hubble Fellow
}

\date{Accepted XXX. Received YYY; in original form ZZZ}

\pubyear{2017}

\begin{document}
\label{firstpage}
\pagerange{\pageref{firstpage}--\pageref{lastpage}}
\maketitle

\begin{abstract}
%Recently, there is a discussion about constructing a new,  large optical-infrared telescope, in which the aperture of the primary mirror is 12m \citep{Su et al. 2016}. Su et al.
Recently, Su et al. (2016) propose an innovative design, referred as the ``SYZ" design, for Chinese new project of a 12m optical-infrared telescope. The  SYZ telescope design  consists of three aspheric mirrors with non-zero power, including a relay mirror below the primary mirror. SYZ design yields a good imaging quality and has a relatively flat field curvature at Nasmyth focus. To evaluate the science-compatibility of this three-mirror telescope, in this paper, we thoroughly compare the performance of SYZ design with that of  Ritchey-Chr\'etien (RC) design, a conventional two-mirror telescope design. Further, we propose the Observing Information Throughput ($OIT$) as a  metric for quantitatively evaluating the telescopes' science performance. We find that although a SYZ telescope yields a superb imaging quality over a large field of view, a two-mirror (RC) telescope design holds a higher overall throughput, a better diffraction-limited imaging quality in the central field of view (FOV$<5'$) which is better for the performance of extreme Adaptive Optics (AO), and a generally better scientific performance with a higher $OIT$ value.
\end{abstract}

\begin{keywords}
astronomical optics -- telescopes -- image quality assessment -- performance metric
\end{keywords}

%%%%%%%%%%%%%%%%%%%%%%%%%%%%%%%%%%%%%%%%%%%%%%%%%%

%%%%%%%%%%%%%%%%% BODY OF PAPER %%%%%%%%%%%%%%%%%%

\section{Introduction}
\begin{figure}
	\includegraphics[width=\columnwidth]{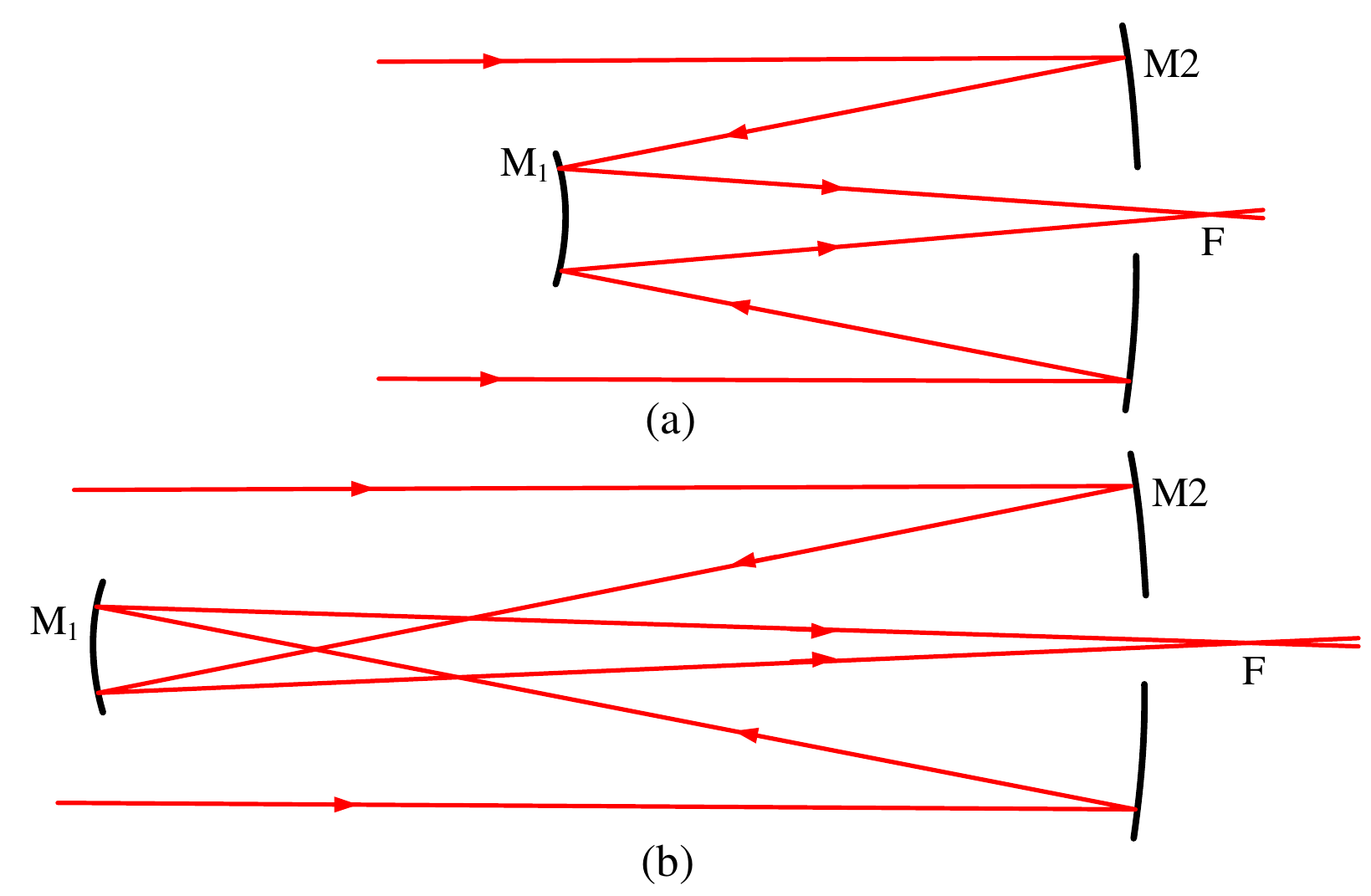}
    \caption{Typtical two-mirror designs: (a) RC design; (b) AG design.}
    \label{fig:fig1}
\end{figure}

The development of new technologies yields increasingly advanced astronomical telescopes with novel instruments. These new technologies constantly improve the telescope's aperture, throughput, angular resolution, spectral resolution, and time resolution of astronomical observations (\citealt{Nelson et al. 1985}; \citealt{Nelson 2000}). Besides, researchers have also explored a variety of configurations for the telescope designs due to their specific applications and requirements. Among them, the two-mirror designs include the Ritchey-Chr\'etien (RC) configuration as shown in Figure~\ref{fig:fig1}(a) and Aplanatic Gregorian (AG) configuration as shown in Figure~\ref{fig:fig1}(b), are the most prevalently adopted designs for current large telescopes. These RC design telescopes include 8m Subaru Telescope, Keck (10m) telescope, and the next generation Thirty-Mirror-Telescope (TMT) that is currently under construction (\citealt{Nelson et al. 1985}; \citealt{Nelson 2000};  \citealt{Kaifu 1998}). The most typical examples for Aplanatic Gregorian configuration are the Large Binocular Telescope (LBT) (8.4m$\times$2) (\citealt{Hill et al. 2003}) and the next generation Giant Megallanic Telescope (GMT) (\citealt{Johns 2008}). The RC and AG configurations are concise and contain only two aspheric mirrors with non-zero power: the primary mirror and the secondary mirror.  Another plane tertiary mirror with no power is required to direct the incident light beam into the Nasmyth instrument platforms for both RC and AG designs.

Recently, there is a new project of constructing a large optical-infrared telescope (LOT) in China. The aperture of primary mirror is 12m. \citet{Su et al. 2016} propose an innovative design that consists of three aspheric mirrors with non-zero power, including a primary mirror, a secondary mirror, and a relay mirror (referred as SYZ relay mirror) just below the primary mirror. The configuration of the conceptual design along with the comprehensive parameters can be found in \citet{Su et al. 2016}. %For simplicity, we can take the innovative design as one typical example of a 3-mirror configuration, since it contains three aspheric mirrors with non-zero power along the complete optical path for Nasmyth focus. 
According to their statements, the SYZ design yields excellent geometrical image quality over a large field of view (FOV) compared to the traditional 2-mirror design and relatively more flat field curvature at Nasmyth focus. Further, SYZ deisgn can switch between Coude focus and Nasmyth focus conveniently (\citealt{Su et al. 2016}). 

To better evaluate the effectiveness of this novel SYZ design, we compare the performance of the SYZ design with that of conventional two-mirror design (e.g., RC design) based on several different performance metrics. Several independent metrics have already been proposed to  evaluate the telescope designs, including the calculation of the throughput and the imaging quality metrics (e.g., wavefront error; 80\% enclosed energy diameter; equivalent noise area (\citealt{King 1983}); normalized point source sensitivity (\citealt{Seo et al. 2009})). These quantities help to understand the different designs and better budget errors of telescopes. In this paper, we use these metrics and also introduce new metrics to directly and quantitatively compare the scientific productivity of the SYZ design with that of the RC design. This paper is organized as follows: in \S2 and \S3, we present the design of a RC and a SYZ system. In \S4, we first analyze the diffraction image quality and system's throughput of all designs. Then we compare the telescope's effective apertures under seeing limited observation and ground layer adaptive optics (GLAO) corrected observation. Finally, we propose a new performance metric of telescope named as observational information throughput ($OIT$), which is directly related to the scientific productivity of telescopes with different designs in this paper. Since the Nasmyth focus hold a majority of instruments for current large telescopes, all the performance metrics are principally measured at Nasmyth focus in the following of this paper.

\section{SYZ DESIGN AND PERFORMANCE}
\subsection{SYZ design}
\begin{figure}
	\includegraphics[width=\columnwidth]{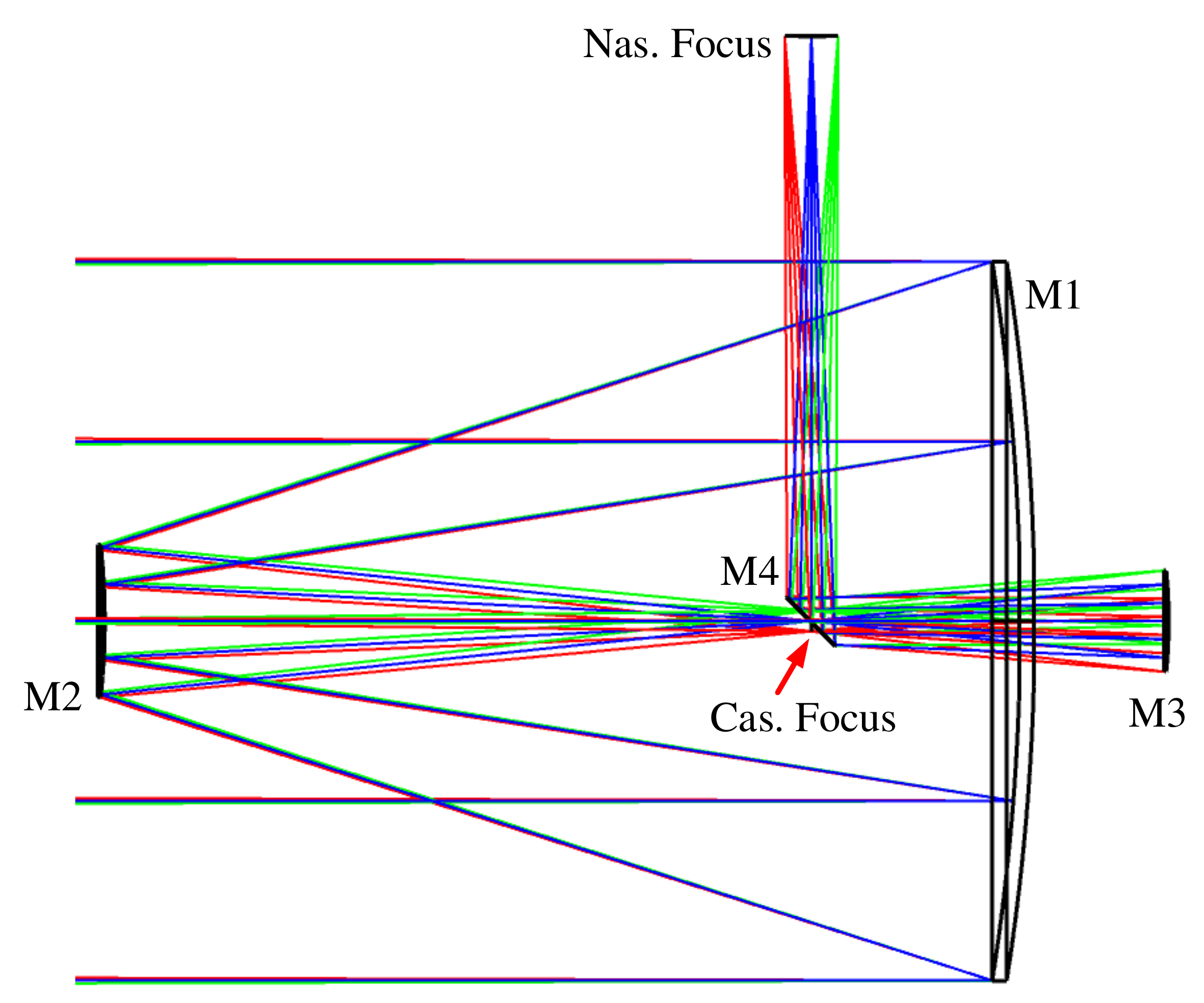}
    \caption{The Nasmyth system of 12-m SYZ telescope (M1: primary mirror; M2: secondary mirror; M3: SYZ relay mirror; M4: flat fold  mirror). Different color represents the different field radius. The field radius is defined as the angle between incident light and telescope optical axis. }
    \label{fig:fig2}
\end{figure}

This SYZ design was proposed by \citet{Su et al. 2016}. The SYZ design contains three mirrors with non-zero power, i.e. primary mirror, secondary mirror and the SYZ relay mirror, as shown in Figure~\ref{fig:fig2}. It is an example of 3-mirror telescopes. The parameters of different mirrors are listed in Table~\ref{tab:TableOne}, and the focal ratio at Nasmyth focus is set to be f/12.8. The primary mirror (M1) has a hyperbolic surface. Both M2 surface and M3 surface are aspheric surfaces. The effective focal length ($EFFL$) of the system is 153,595mm. The maximum FOV of the system is set as 20 arcmin. In addition, the f-number of the primary mirror is set to be 1.6.

% Table 1
\begin{table*}
	\centering
	\caption{The design parameters of the SYZ system.}
	\label{tab:TableOne}
	\begin{tabular}{|c|c|c|c|c|c|c|} % four columns, alignment for each
		\hline
		Element & Curvature radius (mm) & Thickness (mm) & Aperture diameter (mm) & Conic & a6 & a8 \\
		\hline
		M1 & -38400 & -15297.69 & 12000 & -0.960336 & 4 & 4\\
		M2 & -11657.77 & 17742.68 & 2500 & -3.368228 & -4.594e-23 & 1.556e-30\\
		M3 & -8164.38 & -5936.13 & 1458 & -0.638467 & -3.962e-22 & 1.942e-28\\
        M4 & $\infty$  & 9760 & 1133x787 & - & - & -\\
		Image & 8851.47 & - & 894 & - & - & -\\
		\hline
	\end{tabular}
\end{table*}

\subsection{System performance at Nasmyth (Nas.) focus}
\label{sec:SYZ performance} % used for referring to this section from elsewhere

\begin{figure}
	\includegraphics[width=\columnwidth]{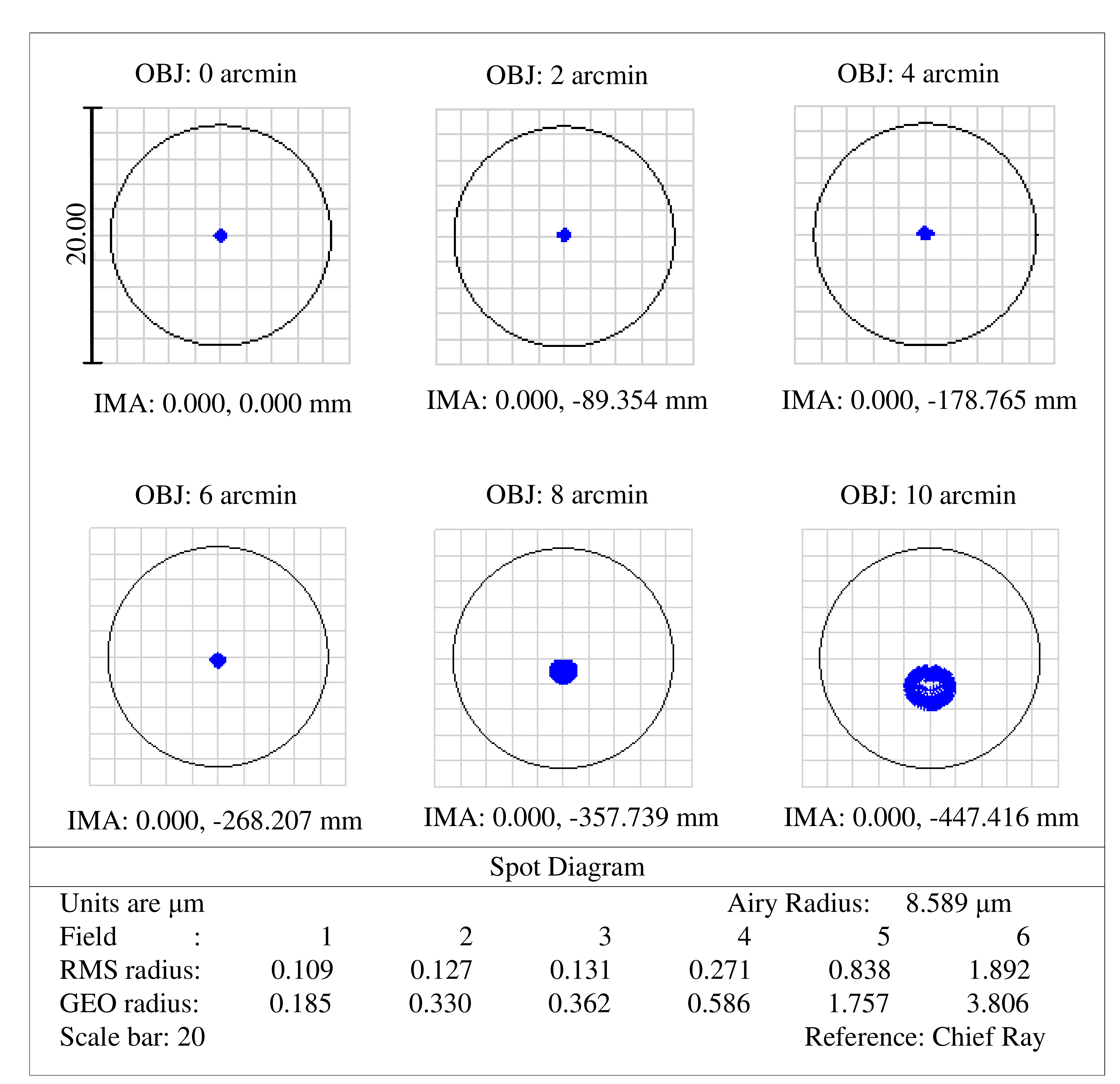}
    \caption{Geometrical spot diagrams of different FOVs at Nasmyth focus of SYZ telescope.}
    \label{fig:fig3}
\end{figure}

\begin{figure}
	\includegraphics[width=\columnwidth]{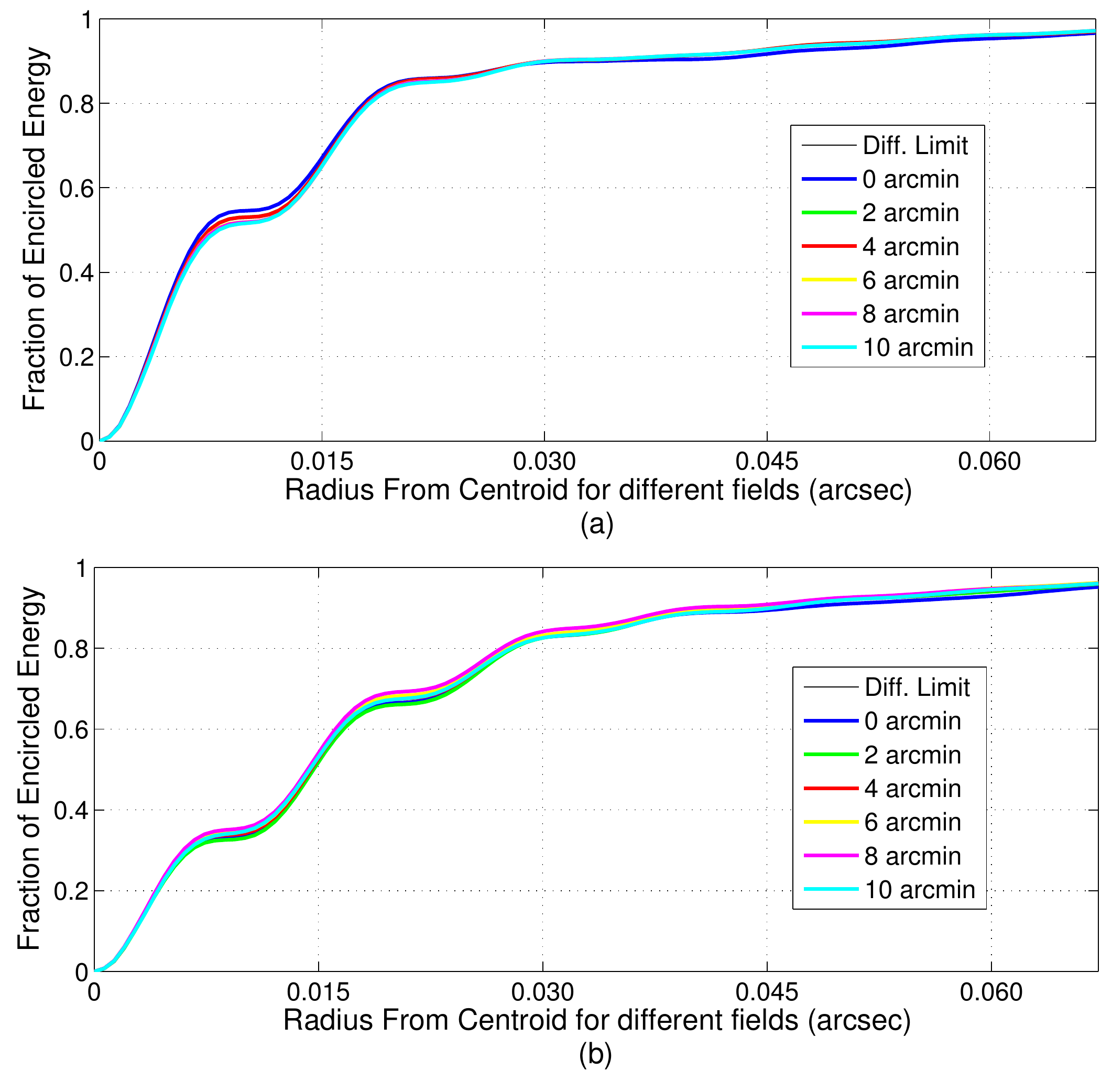}
    \caption{The fraction of encircled energy as a function of image spot radius in the diffraction limited scenario: (a) Central obscuration ratio ($\epsilon$) = 0.35; (b) Central obscuration ratio ($\epsilon$) = 0.5. The different colors represent the different field radius. The field radius is defined as the angle between the incident light and the telescope axis.}
    \label{fig:fig4}
\end{figure}

For the imaging quality of SYZ design, Figure~\ref{fig:fig3} shows the spot diagrams at different field positions. The fraction of encircled energy as a function of image spot radius is shown in Figure~\ref{fig:fig4}. From these results, we can see that the SYZ system has an exceptional geometrical image quality: the diameter of 80\% encircled energy (EE80) less than 0.002 arcsec for the full FOV. Besides, the curvature radius of the Nasmyth focal surface of SYZ system is about 8.85m, which is relatively flat and thus easier for implementation of instruments with large FOV. These results are all consistent with that reported in \citet{Su et al. 2016}. However, it is well-known that the geometrical image spot size cannot fully reflect the performance of a diffraction-limited optical system since the diffraction effect will dominate the final image quality.. For the wavelength of 0.55$\mu$m, the Airy disk's full width at half-maximum ({\it FWHM}) diameter for diffraction-limited condition %without obscuration 
is specified as (\citealt{Airy 1835}): 

\begin{equation}
    \theta _{\rm{Airy}}=1.22\frac{\lambda _{0.55}}{D},
	\label{eq:airy}
\end{equation}

\noindent
where $\lambda_{0.55}$ is the wavelength at 550nm, $D$ is the effective entrance aperture diameter of the optical system. The ideal Airy disk size of the LOT optical system with an aperture of 12 m is 0.0115'' by Equation~(\ref{eq:airy}). A linear relation between EE80 and {\it FWHM}: EE80=$1.6\times FWHM$. Thus, the EE80 diameter of ideal Airy disk for LOT telescope is 0.018 arcsec, which is much larger than the geometric image spot size. For the diffraction-limited scenario, diffraction dominates the final image quality, and the actual image spot size can be approximately derived by:

\begin{equation}
    \theta _{\rm{Spot}}=\sqrt{\theta _{\rm{Airy}}^{2}+\theta _{\rm{Geo}}^{2}}.
	\label{eq:Spot}
\end{equation}

\noindent
Since $\theta _{\rm{Geo}}$ is ignorable (less than 0.002'' in full FOV) in the SYZ design, the size of SYZ spot size in the diffraction-limited scenario is close to that of the Airy disk. 

In order to further evaluate the image quality of SYZ telescope, we used ZEMAX to take a quantitative evaluation of image spots' encircled energy (EE) distribution for SYZ systems with different central obscuration ratios. The central obscuration ratio is defined as the relative linear size of the M4's central hole to the M4's linear size. We denote this central obscuration ratio as $\epsilon$, which determines the FOV at Nasmyth focus. We present the EE as a function of image spot radius in Figure 4:  Figure 4 (a) showing EE vs. radius with $\epsilon=0.35$, corresponding to a FOV of 14 arcmin; and Figure 4 (b) showing EE vs. radius with $\epsilon=0.5$, corresponding to a FOV of 20 arcmin.  The 80\% enclosed energy (EE80) is about 0.036 arcsec when the relative size of M4's central hole is $\epsilon= 0.35$; and the EE80 is about 0.055 arcsec when $\epsilon=0.5$.

Based on our calculations in Appendix~\ref{sec:COdiff} for the obscured diffraction pattern (e.g., \citealt{Airy 1835}; \citealt{Sacek 2006}), the EE80 diameter of image spot with central obscuration of 0.5 (0.25 for the area obscuration) is more than twice of the Airy disk size, consistent with our previous ZEMAX simulation results. We conclude that the SYZ's on-axis resolution is relatively lower than the two mirror systems. Further, the larger size of M4's hole for the SYZ system will result in a worse on-axis image quality.. In next section, we will provide more discussions about the effect of M4's central hole such as central obstruction, FOV vignetting, and extra diffraction degradation.

\subsection{SYZ system performance at Cassegrain (Cas.) focus}

SYZ is optimally designed for Nasmyth focus, and thus the image quality at Cassegrain focus is relatively worse because of lacking symmetry between the two foci. In the current SYZ design (\citealt{Su et al. 2016}), the EE80 diameter of in the central FOV  is about 1.87 arcsec, which can be reduced to 1.0 arcsec with further optimization at Cassegrain focus by increasing the telescope tube length (equivalently increasing the Cassegrain focus ratio). Considering the image quality issue as well as other instrumentation problems (such as location of this focus), the SYZ design will be difficult to be compatible with Cassegrain focus. However, Cas. focus is important for some two-mirror telescopes because of its superb throughput and symmetric optics. For example, Keck telescope supports LRIS, ESI and MOSFIRE (\citealt{Oke et al. 1995}; \citealt{Sheinis et al. 2002}; \citealt{McLean et al. 2012}) at Cas. focus. Lacking a good Cas. focus may be a potential drawback for SYZ design. 

\section{RC DESIGN AND PERFORMANCE}

\begin{figure}
	\includegraphics[width=\columnwidth]{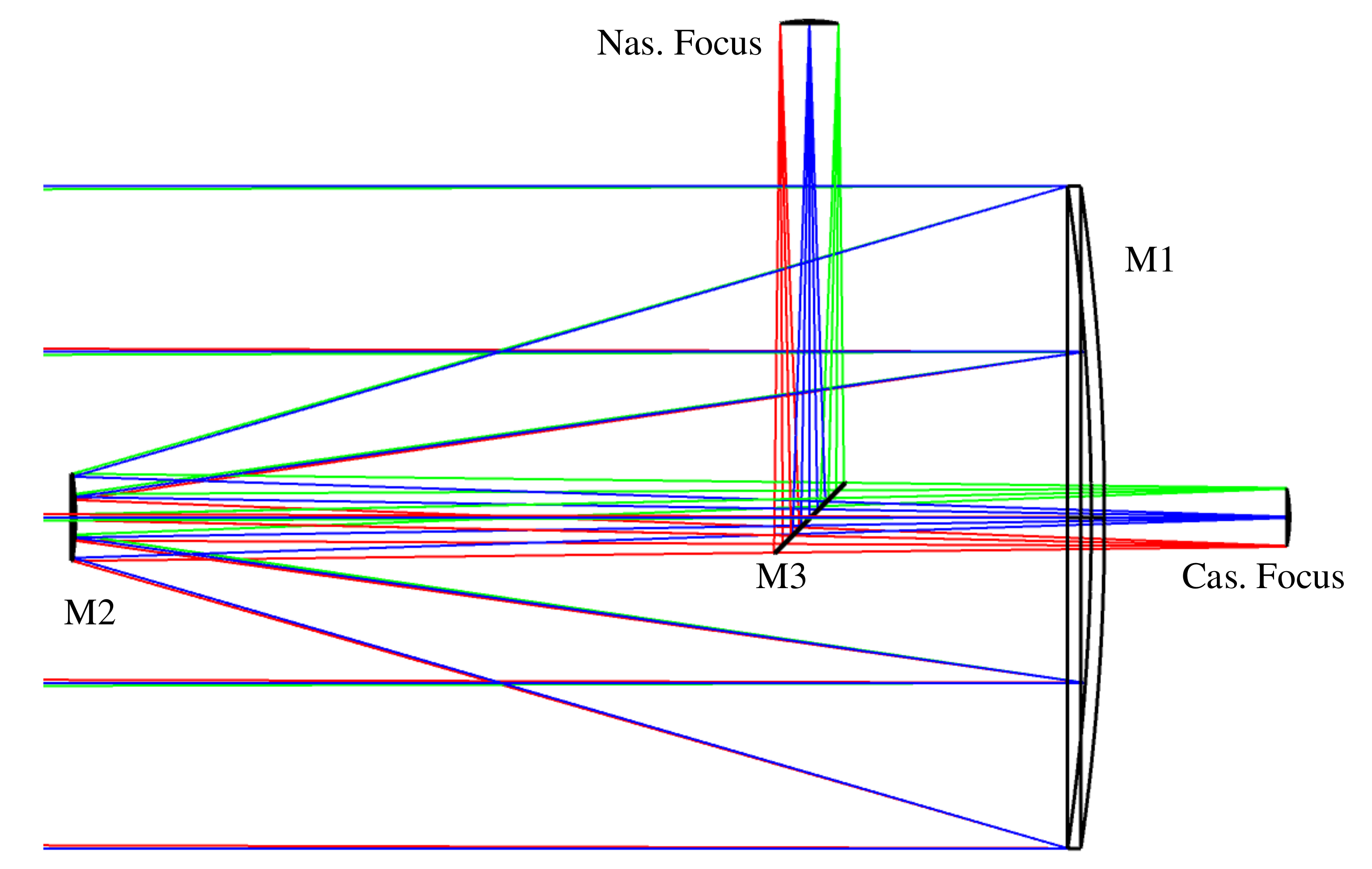}
    \caption{The 2D layout of a RC telescope design. Different colors represent the different field radius.}
    \label{fig:fig5}
\end{figure}

As we have stated above, the most widely used designs for large telescopes are two-mirror designs, such as RC and AG systems. Compared to a three-mirror system, a two-mirror telescope has simpler structure which probably yields less construction cost. %In this section, we will focus on analyzing properties of Ritchey-Chr\'etien telescope. 
In this section, we design a Ritchey-Chr\'etien telescope. The design parameters are listed in Table~\ref{tab:TableTwo} and the 2D layout of the RC telescope is shown in Figure~\ref{fig:fig5}. We set both Nasmyth focus and Cassegrain focus at f/15 and thus the effective focal length of the system is 180m. The f-number of the primary mirror is set as 1.60, which is the same as current SYZ design. The maximum FOV of RC design is set to be 20 arcmin.

% Table 2
\begin{table*}
	\centering
	\caption{The design parameters of SYZ configuration at Nasmyth focus}
	\label{tab:TableTwo}
	\begin{tabular}{|c|c|c|c|c|} % four columns, alignment for each
		\hline
		Element & Curvature radius (mm) & Thickness (mm) & Aperture diameter (mm) & Conic \\
		\hline
		M1 & -38400 & -1.696E+4 & 12000 & -1.0030048 \\
		M2 & -5014.97 & 12000 & 1540 & -1.5707753  \\
        M3 & $\infty$  & 9000 & 1740x1230 & -\\
		Image & 2641.18 & - & 1045 & - \\
		\hline
	\end{tabular}
\end{table*}

\begin{figure}
	\includegraphics[width=\columnwidth]{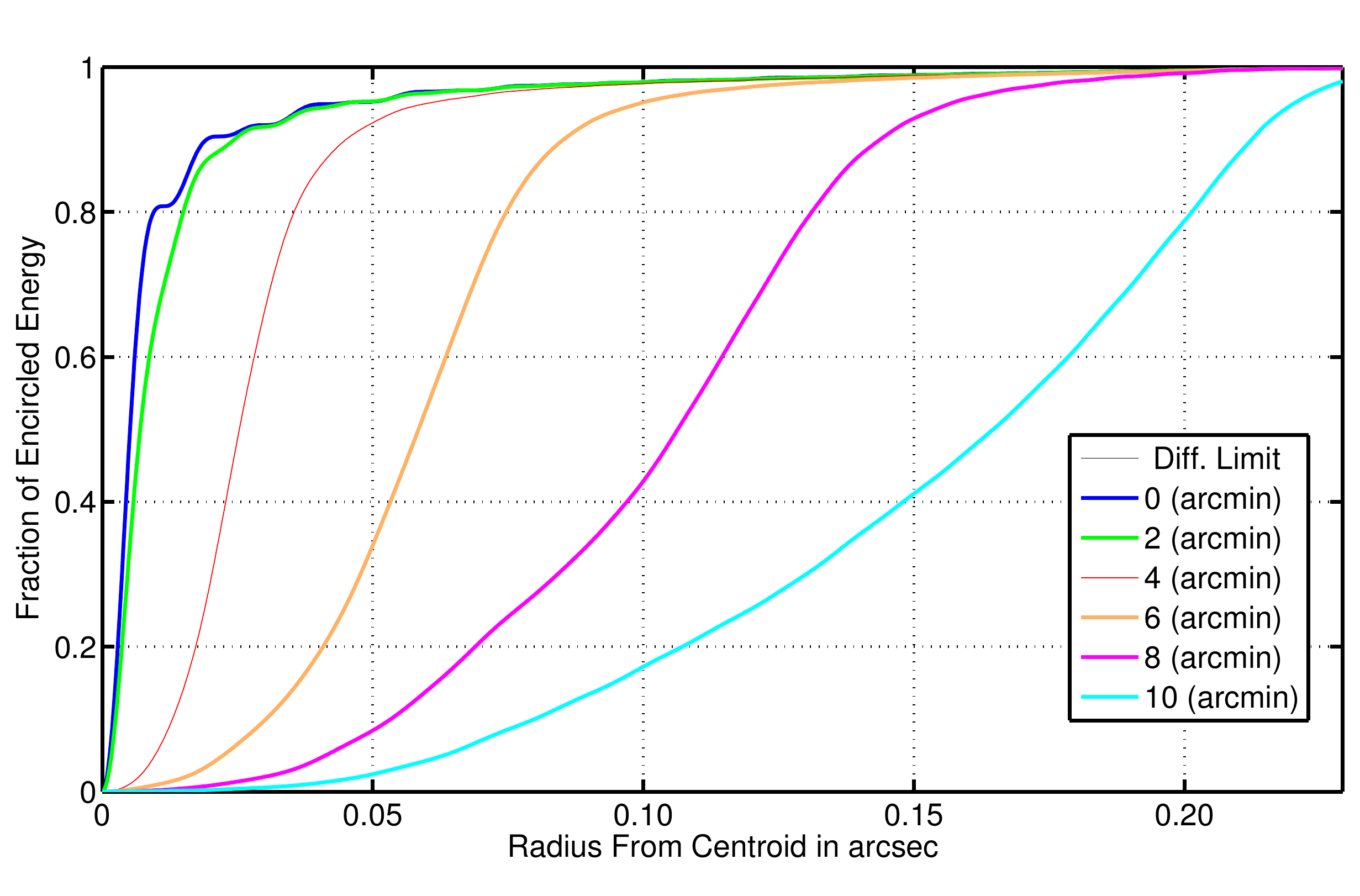}
    \caption{The fraction of encircled energy vs. radius from centroid of image spots with diffraction effect for RC design. Different colors represent the different field radius. }
    \label{fig:fig6}
\end{figure}

In Figure~\ref{fig:fig6}, we present the fractional encircled energy distribution of image spots with diffraction effect for the RC design. Obviously, image spots at different field positions have very different sizes, which, however, are relatively uniform for the SYZ telescope design. The diameters of image spots with diffraction (EE80) for RC design vary from 0.0193 arcsec (central field position) to 0.3853 arcsec (marginal field position). In addition, the curvature radius of the focal surface for RC design is estimated to be 2.64m, which is less flat compared to that of the SYZ telescope.

\section{COMPARISON OF SYSTEM PERFORMANCE BETWEEN THE SYZ DESIGN AND RC DESIGN}

Compared to two-mirror telescope systems, the SYZ telescope has relatively small field curvature and excellent geometrical image quality for full FOV. However, the SYZ system has a more complex telescope structure, higher central obstruction ratio, higher reflection loss introduced by an additional mirror, and worse on-axis image quality caused by annular pupil diffraction effect due to the central obstruction. In the following sections, we will provide a systematic comparison between RC design and SYZ design. We will focus on the diffraction image quality, system throughputs, and the actual scientific performance. We will apply several telescope evaluation criteria, such as EE80, effective aperture, central intensity ratio ($CIR$), equivalent noise area ($ENA$), and observing information throughput ($OIT$) to evaluate the performance of various telescopes.

\subsection{Diffraction image quality}
\label{sec:diffraction}

Again, we use EE80 diameter in diffraction limited scenario to evaluate the image qualities of different designs. In Table~\ref{tab:TableThree}, we show the comparison results of image spot sizes at Nasmyth focus for different telescope design, where $\epsilon$ specifies the relative size of central obscuration of M4 in SYZ design.

% Table 3
\begin{table}
	\centering
	\caption{The comparison of SYZ designs and RC design in EE80 diameter. The unit of EE80 is in arcsec.}
	\label{tab:TableThree}
	\begin{tabular}{|c|c|c|c|} % four columns, alignment for each
		\hline
		%HFOV (arcmin) & EE80 Diameter (arcsec)\\
        \multirow{2}{1.5cm}{HFOV (arcmin)} & \multicolumn{3}{p{5cm}|}{\footnotesize{Diffraction EE80 (Diameter (arcsec))}}  \\
         \cline{2-4} & SYZ($\epsilon$=0.5) & SYZ($\epsilon$=03.5) & RC \\
		\hline
		0 & 0.0544 & 0.0358 & 0.0193 \\
		2 & 0.0546 & 0.0360 & 0.0297 \\
        4 & 0.0548 & 0.0362 & 0.0677 \\
		6 & 0.0550 & 0.0364 & 0.1420 \\
        8 & 0.0552 & - & 0.2489 \\
		6 & 0.0554 & - & 0.3583 \\
		\hline
	\end{tabular}
\end{table}

From the comparison results, the size of M4's central hole plays a key role for optical system's performance of SYZ designs: on one hand, it degrades the image quality due to its diffraction effect; on the other hand, vignetting effect caused by the central hole occurs when FOV increases. As we have stated above, if we need unvignetted full FOV of 14 arcmin, then the relative size of the central hole of M4 must be larger than 0.35. The relative size of M4's hole is 0.5 for an unvignetted FOV of 20 arcmin. However, the larger the central hole of M4, the more degradation it will bring to the image quality. Due to the diffraction effect of M4's central hole, the image quality of central FOV (within 5 arcmin for $\epsilon=0.35$; within 7 arcmin for $\epsilon=0.5$) for SYZ designs is not as good as that for the two-mirror designs. However, we can reach diffraction-limited observation with the help of adaptive optics technologies such as extreme adaptive optics (ExAO) and multi-conjugate adaptive optics (MCAO), whose FOVs are currently restricted to less than 1 arcmin (\citet{Macintosh et al. 2006}; \citet{Rigaut et al. 2000}). For GLAO, although its FOV might reach 10 arcmin, it can only improve image quality by a factor of $\sim$2,  far from diffraction limited,  and its performance strongly relies on turbulence profile  (\citealt{Hart et al. 2010}; \citealt{Andersen et al. 2006}; \citealt{Orban de Xivry et al. 2016}). For the image quality at large field radius, the atmosphere seeing (median 0.7-- 0.8 arcsec in several candidate sites for this project) and telescope instrumentation errors will dominate over full FOV of 15 arcmin at Nasmyth or Cassegrain focus (\citealt{Nelson et al. 1985}). Thus, we conclude as follows. First, the image quality of the three-mirror designs is worse than that of the two-mirror designs at the central field, especially within central FOV of 1 arcmin. Second, at the edge of large FOV (e.g., 15 -- 20 arcmin), performance would not differ much between two-mirror designs and three-mirror designs due to seeing limited condition. Third, if we further consider the effect of instrumentation error, the image quality of SYZ design is likely worse than theoretical expectation (e.g., Figure 4) due to its more complex alignment. 

\subsection{Throughput analysis}
\label{sec:throughput}

To maximize the throughputs of the telescope optical systems, we need to apply reflective metal coating to all mirror surfaces to increase the reflectivity. Generally, aluminum, silver and gold are the three most common metal coatings (\citealt{Bennett et al. 1965}). Aluminum is the most popular coating material for most of the astronomical observatories. Aluminum is reflective over the full range of wavelengths from 300nm to 25$\mu$m. However, the reflectivity of aluminum coating in the 300-1000 nm range is only about 90\%. Silver is a better coating choice for wavelength that is longer than 340nm, but its oxidization rate is fast, so that its reflectivity drops rapidly if coated naked. It also has very low reflectivity for wavelength shorter than 340nm.  The gold coating suffers from the same problem as the silver coating and it has an even lower ultra-violet reflectivity (\citealt{Bennett et al. 1965}).

For Aluminum coating, we assume that its lifetime is 2 years. The reflectivity of aluminum coating in the optical wavelength range will degrade from 90\% to about 87\% after 1 year, and to 84\% after two years (\citealt{Magrath 1997}). For simplicity, we take the reflectivity of aluminum coating for the full optical wavelength range as 87\% during a 2-year lifetime. However, if we do not consider the ultraviolet band and adopt the enhanced silver coating, then the average reflectivity in optical wavelength range can reach as high as 95\% (\citealt{Vucina et al. 2006}). That means if we use Al coating, then an extra mirror of SYZ telescope yields 13\% extra light loss. If one adopts enhanced silver coating, then an extra mirror of SYZ yields 5\% extra light loss. 

As stated above, another important light loss is from the central obscuration effect caused by the optical elements along the optical path. In this paragraph, we quantitatively evaluate this effect. For the SYZ design, secondary mirror, central hole of primary mirror, and central hole of M4 will obscure light near the optical axis. Moreover, the obscuration ratio for the three elements are 0.212, 0.117, and 0.35 (FoV=15') or 0.50 (FoV = 20') respectively. Typically, the central hole of M4 in either design dominates the central obscuration effect along the optical path for Nasmyth focus of SYZ design. For RC design, only the secondary mirror has a central obstruction effect along the optical path. Thus, considering reflection loss and central obstruction effect, the final throughput of all designs can be denoted by

\begin{equation}
    \eta =r_{C}^{N}\cdot \left ( 1-\epsilon _{M}^{2} \right ),
	\label{eq:throughput}
\end{equation}

\noindent
where $r_{C}$ is the reflectivity of mirror coating, $N$ is the number of mirrors along the optical path, and $\epsilon _{M}$ is the central obstruction ratio of telescope system, which is generally decided by element with maximum central obscuration along the optical path.  Based on Equation~(\ref{eq:throughput}), the total throughputs of all designs can be concluded in Table~\ref{tab:TableFour}.

% Table 4
\begin{table*}
	\centering
	\caption{The throughput of SYZ designs and RC designs}
	\label{tab:TableFour}
	\begin{tabular}{|c|c|c|c|c|} % four columns, alignment for each
		\hline
		%Design & CO & Total throughput\\
        \multirow{2}{1.5cm}{Design}& \multirow{2}{1.5cm}{Central Obscuration}& \multicolumn{3}{p{5cm}|}{\centering Total Throughput} \\
         \cline{3-5} & & New Al & Two-year Al & Enhanced Ag \\
		\hline
		SYZ($\epsilon=0.35$) & 0.35 & 57.6\% & 50.3\% & 71.5\% \\
		SYZ($\epsilon=0.50$) & 0.50 & 49.2\% & 43.0\% & 61.1\% \\
        RC(Nas. focus) & 0.13 & 71.8\% & 64.9\% & 84.4\% \\
		RC(Cas. focus) & 0.13 & 79.7\% & 74.5\% & 88.8\% \\
		\hline
	\end{tabular}
\end{table*}

\subsection{Effective aperture estimation}
\label{sec:aperture}

The SYZ design has a good performance in image quality at large FOV, but its image quality at central field position (FoV $<5'$) and the system throughput are not as good as that of the RC design. The effective aperture is generally used as a parameter to evaluate the signal-to-noise ratio (SNR) under a given exposure time ($t _{\rm{exp}}$). In the following, we will take the effective aperture as one of the general criteria to evaluate the optical performance of both RC and SYZ systems at different FOVs. Nowadays, most astronomical observations are conducted in so-called "background-limited" condition, in which sky background is much brighter than the source. In background-limited scenario, the SNR is proportional to the following quantity (\citealt{Budding et al. 2007}): 

%Both throughput and imaging quality can be reflected in the "effective aperture". 
\begin{equation}
SNR\propto D_{0} \cdot \sqrt{t_{\rm{exp}}}\cdot \sqrt{\eta } \cdot \frac{1}{FWHM},
\label{eq:SNRA}
\end{equation}

\noindent
where $t _{\rm{exp}}$ is the exposure time, and {\it FWHM} is the full width at half maximum of the point spread function (PSF) at the focal surface. Let us assume $D_{0}$ is the aperture of the primary mirror, and then the effective aperture for any optical system with throughput of $\eta$ and $FWHM$ can be denoted as: 

\begin{equation}
    D_{\rm{eff}}\equiv D_{0}\cdot \sqrt{\eta }\cdot \frac{FWHM_{0}}{FWHM},
	\label{eq:DEFF}
\end{equation}

\noindent
where $FWHM _{0}$ is the atmosphere seeing. The definition is provided in such a way so that $D _{\rm{eff}}$ can be used to directly reveal the observed SNR given a $t _{\rm{exp}}$. As we have asserted in Section~\ref{sec:diffraction}, there is little difference between $FWHM$ and $FWHM _{0}$ under seeing-limited condition or GLAO condition. 

Based on Equation~(\ref{eq:DEFF}), we can summarize the effective apertures at $\lambda \sim  0.55\mu m$ of different designs in Table 5 under seeing limited observations or GLAO corrected observations, assuming a typical seeing of $0.7"$ and a factor of $2\times$ improvement of the FWHM ($0.35"$) under the GLAO condition over a 10' FOV (e.g., Anderson et al. 2006). Among all design configurations, SYZ designs have much smaller effective apertures compared to the RC design, and the RC design at Cassegrain focus can have the best SNR for observing faint sources. 

% Table 5
\begin{table}
	\centering
	\caption{The effective aperture of SYZ design and RC design}
	\label{tab:TableFive}
	\begin{tabular}{|c|c|c|c|} % four columns, alignment for each
		\hline
		%Design & EA (m)\\
        \multirow{2}{1.5cm}{Design} & \multicolumn{3}{p{5cm}|}{\centering Effective Aperture (m)} \\ 
      \cline{2-4}  &  New Al & Two-year Al & Enhanced Ag \\
		\hline
		SYZ($\epsilon=0.35$)  & 9.1 & 8.5 & 10.1 \\
		SYZ($\epsilon=0.50$)  & 8.4 & 7.9 & 9.4 \\
        RC(Nas. focus)  & 10.2 & 9.7 & 11.0 \\
		RC(Cas. focus)  & 10.7 & 10.4 & 11.3 \\
		\hline
	\end{tabular}
\end{table}

\subsection{Central intensity ratio (CIR) comparison}
\label{sec:CIR}

In section~\ref{sec:aperture}, we have compared the optical performance of RC design and SYZ design under seeing limited observation and GLAO observations. Most large telescopes nowadays aim to achieve diffraction-limited performance up to a FOV of a few tens of arcsec with extreme adaptive optics (ExAO). In this case, we generally use central intensity ratio ($CIR$) to evaluate the telescope performance, which correlates system's Strehl ratio  and photon throughput (\citealt{Dierickx 1992}).

In the RC or SYZ design, the Strehl ratio degradation of PSF at the center of FOV is largely dependent on the diffraction effect of central obstruction. Based on our derivation of central obscuration (CO) diffraction pattern in Appendix~\ref{sec:COdiff}, one can see that Strehl ratio of a telescope with central obscuration ratio of $\epsilon$ can be estimated as $(1-\epsilon ^{2}) ^{2}$. Considering the reflection loss, telescope's $CIR$ at the center of FOV under ExAO can be expressed by:

\begin{equation}
    CIR=r_{C}^{N}\cdot \left ( 1-\epsilon _{M}^{2} \right )^{2}.
	\label{eq:CIR}
\end{equation}

Based on Equation~(\ref{eq:CIR}), comparison of RC design and SYZ design under ExAO corrected observations can be summarized in Table~\ref{tab:TableSix}. Following the comparison results, one can see that RC design generally has a higher $CIR$ and thus has a better ExAO performance. 

% Table 6
\begin{table}
	\centering
	\caption{The central intensity ratio ($CIR$) of SYZ design and RC design under ExAO}
	\label{tab:TableSix}
	\begin{tabular}{|c|c|c|c|} % four columns, alignment for each
		\hline
		%Design & CIR\\
        \multirow{2}{1.5cm}{Design} & \multicolumn{3}{p{5cm}|}{\centering Central Intensity Ratio } \\ 
      \cline{2-4}  &  New Al & Two-year Al & Enhanced Ag \\
		\hline
		SYZ($\epsilon=0.35$)  & 0.51 & 0.44 & 0.63 \\
		SYZ($\epsilon=0.50$)  & 0.37 & 0.32 & 0.46 \\
        RC(Nas. focus)  & 0.70 & 0.64 & 0.83 \\
		RC(Cas. focus)  & 0.78 & 0.73 & 0.87 \\
		\hline
	\end{tabular}
\end{table}

\subsection{Observing information throughput (OIT) evaluation}
\label{sec:OIT}
In this section, we introduce a new evaluation criterion, the observing information throughput ($OIT$), to evaluate telescope's observing capability for full FOV. Before doing this, we first introduce another general telescope's performance metric, i.e. equivalent noise area ($ENA$). $ENA$ has been first proposed in \citet{King 1983}, and further adopted as a crucial performance metric for large telescopes (\citealt{Nelson et al. 2008}; \citealt{Angeli et al. 2011}). The $ENA$ has a physical meaning of the smallest aperture that can be used to extract all the information of a faint source, which can be defined as:

\begin{equation}
    ENA=\frac{1}{ \iint_{ }^{ }\phi ^{2}dxdy},
	\label{eq:ENA}
\end{equation}

\noindent
where $\phi$ is the point-spread function (PSF).  Detailed derivations of $ENA$ can be found in Appendix~\ref{sec:ENA}. Typically, $ENA$ is the inverse of the point source sensitivity (PSS), which is another telescope performance metric proposed by \cite{Seo et al. 2009}. Assuming a source with $E$ as the total photons per second, we can define the $SNR$ for the observation of the source as:

\begin{equation}
    \frac{S}{N}=\frac{E\cdot t\cdot \eta }{\sqrt{E\cdot t\cdot \eta+B\cdot t\cdot ENA\cdot \eta }},
	\label{eq:SNR}
\end{equation}

\noindent
where $\eta$ is the throughput of the telescope, $E$ is the stellar intensity in photons, and $B$ is the observed background sky level in photons. For the background limited case, the exposure time can be represented by:

\begin{equation}
    t=\frac{ENA}{\eta }\frac{S}{N}\frac{B}{E^{2}}.
	\label{eq:time}
\end{equation}

Based on Equation~(\ref{eq:time}), we can see that the integration time is proportional to $ENA$ for observing faint sources. As long as the total energy of the source ($E$), and background ($B$) is fixed, the integration time is proportional to the $ENA/\eta$ given the same $S/N$. Thus, $\eta/ENA$ can be used as a direct parameter to evaluate the science productivity given a FOV and a given exposure time. In addition, we consider that the survey speed is proportional to FOV area of telescope. Thus, we suggest the integration of $\eta/ENA$ over the whole FOV of telescope as an independent, dimensionless criterion for any telescope design.  We call this new criterion as observing information throughput ($OIT$), which can be defined as

\begin{equation}
    OIT=\int_{0}^{2\pi }\int_{0}^{FOV/2}\frac{\eta }{ENA}rdrd\theta,
	\label{eq:OIT}
\end{equation}

\noindent
where $r$ is the field radius of telescope measured by radian (or arcmin), and $ENA$ is a quantity to evaluate the image quality on the focal surface of telescope. $OIT$ directly measures the total science output of telescopes with a given exposure time. 

The total $ENA$ of the telescope can be approximated by:

\begin{equation}
    ENA=ENA_{\textup{Seeing}}+ENA_{\textup{Astig}}+ENA_{\textup{Error}},
	\label{eq:ENAT}
\end{equation}

\noindent
where $ENA _{\textup{Seeing}}$ is the $ENA$ caused by atmosphere seeing, $ENA _{\textup{Astig}}$ is the $ENA$ caused by astigmatism of telescope design, and $ENA _{\textup{Error}}$ is the $ENA$ caused by the error due to the telescope instrumentation.  Based on our definition of $ENA$, the value of $ENA _{\textup{Seeing}}$ for atmosphere seeing of 0.7 arcsec is about 1.20 arcsec$^{2}$ (assuming PSF is a Gaussian profile). Since it is hard to quantify the instrumentation error, we assume that we can achieve a perfect instrumentation of telescope for all designs, i.e. $ENA _{\textup{Error}}=0$. The SYZ design is diffraction-limited design for whole FOV, so its $ENA _{\textup{Astig}}$ can be taken as zero too. While for the two-mirror design, $ENA _{\textup{Astig}}$ can be speculated by:

\begin{equation}
    ENA_{\textup{Astig}}=\kappa \cdot AAST^{2},
	\label{eq:ENAA}
\end{equation}

\noindent
where $\kappa$ is a constant with value of $\pi$/4 that can be derived based on the PSF caused by astigmatism and the definition of $ENA$, and $AAST$ is the value of astigmatism for RC design at a given field radius of $r$, which can be expressed by:

\begin{equation}
    AAST=AAST_{15}\cdot \left ( \frac{2r}{15'} \right )^{2},
	\label{eq:AAST}
\end{equation}

\noindent
where $AAST _{15}$ is the astigmatism of RC design at field diameter of 15' and can be estimated as 0.243'' based on our simulation result in ZEMAX. Thus, through a simple calculation, over a field diameter of 15', the total $ENA$ is dominated by atmosphere seeing rather than telescope design aberrations, which once again demonstrates our previous conclusion in Section~\ref{sec:diffraction}.

After figuring out the $ENAs$ of all telescope designs proposed above, we can achieve related $OITs$ for these telescope configurations with different coatings separately. The results are as shown in Table~\ref{tab:TableSeven}. Based on the $OIT$ results, we can see that SYZ designs have a lower $OIT$ value compared to RC design at both Nasmyth focus and Cassegrain focus. 

% Table 7
\begin{table*}
	\centering
	\caption{The Observating Information Throughput (OIT) of the SYZ design and RC design.}
	\label{tab:TableSeven}
	\begin{tabular}{|c|c|c|c|c|c|} % four columns, alignment for each
		\hline
		%Design & FOV (arcmin) & ENA(FOV) (arcsec $^{2}$) & OIT (X10$^{5}$)\\
       \multirow{2}{1.5cm}{Design} & \multirow{2}{1.5cm}{FOV (arcmin)} & \multirow{2}{1.5cm}{ENA (FOV)} & 
       \multicolumn{3}{p{5cm}|}{\centering Observing Information Throughput (OIT) ($\times 10^5$) } \\ 
      \cline{4-6} & &  &  New Al & Two-year Al & Enhanced Ag \\
		\hline
		SYZ($\epsilon=0.35$) & 14' & 1.20 & 2.66 & 2.32 & 3.30 \\
		SYZ($\epsilon=0.50$) & 20' & 1.20 & 4.63 & 4.05 & 5.76 \\
        RC(Nas. focus) & 20' & 1.35 & 6.74 & 6.09 & 7.92 \\
		RC(Cas. focus) & 20' & 1.35 & 7.49 & 7.00 & 8.34 \\
		\hline
	\end{tabular}
\end{table*}

In order to make a systematic comparison between RC design and SYZ design in any given FOV, we further quantify the $OITs$ of all optical systems designed with a wider range of FOVs. For the RC design, the central obstruction ratio caused by the secondary mirror for zero FOV is 0.117, which is 0.128 for 20' FOV.  There is a linear relationship between the required size of secondary mirror and the designed maximum FOV, and the central obstruction ratio caused by secondary mirror in a RC system designed for a given FOV can be derived by: 

\begin{equation}
    \epsilon _{M2-RC}=0.117+0.011\cdot \frac{FOV}{20'}.
	\label{eq:RCCO}
\end{equation}

\noindent
While for the SYZ design, the central obscuration ratio caused by the secondary mirror for zero FOV is 0.205, which is 0.213 for 20' FOV; the central obscuration ratio caused by M4 for a given FOV is proportional to the size of M4's central hole, which is also directly proportional to the designed FOV of SYZ telescope. Thus, the central obscuration of the SYZ system at a given FOV can follow the following expressions:

\begin{equation}
    \left\{\begin{matrix}
\epsilon _{M2-SYZ}=0.205+0.008\cdot \frac{FOV}{20'}\\ 
\epsilon _{M4-SYZ}=0.35\frac{FOV}{14'}\\ 
\epsilon _{SYZ}=max\left ( \epsilon _{M2-SYZ}, \epsilon _{M4-SYZ} \right )
\end{matrix}\right.
	\label{eq:SYZCO}
\end{equation}

\begin{figure}
	\includegraphics[width=\columnwidth]{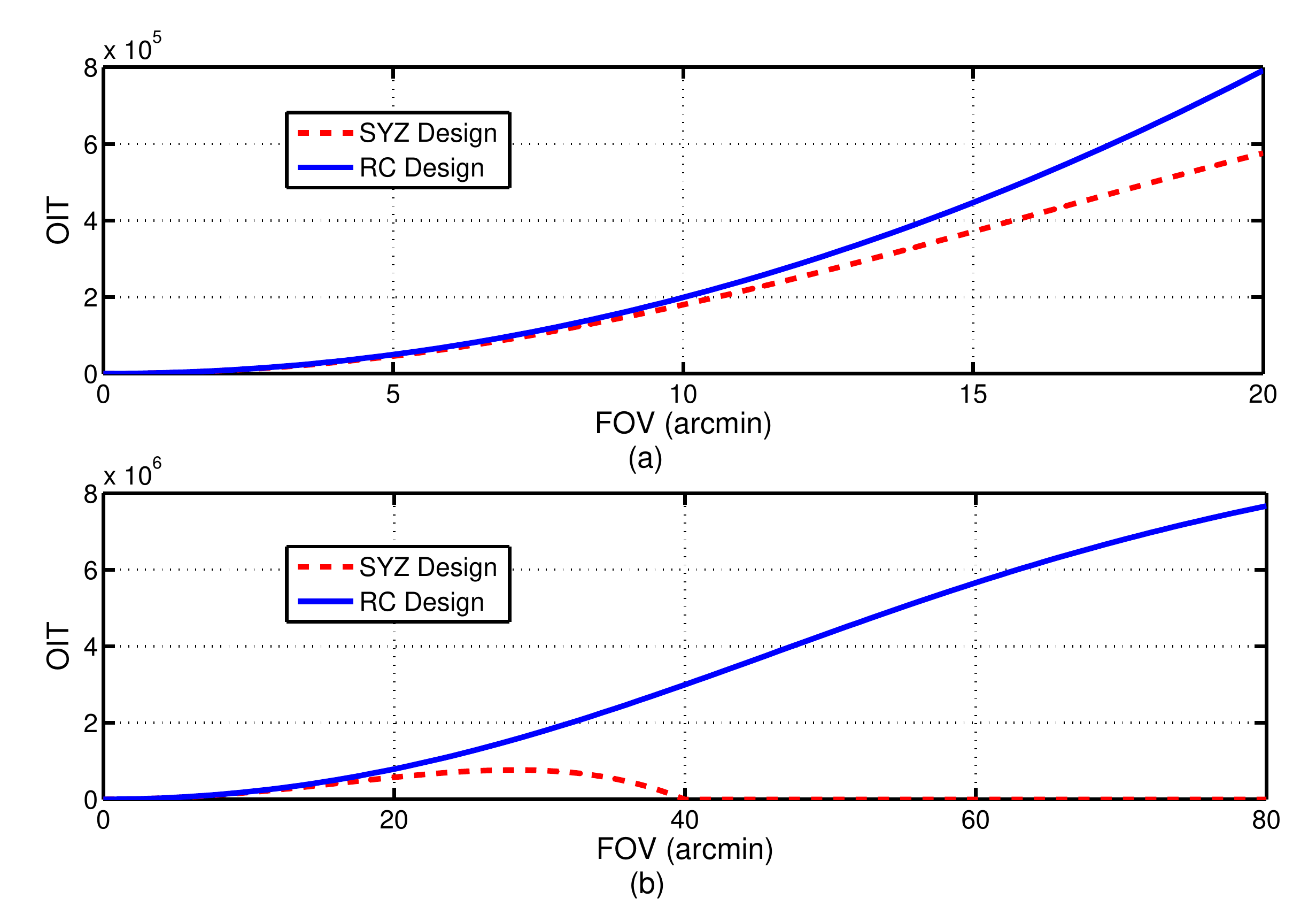}
    \caption{The observing information throughput ($OIT$) of both SYZ and RC configurations as a function of FOV at Nasmyth focus: (a) limiting maximum FOV to 20 arcmin; (b) limiting maximum FOV to 80 arcmin.}
    \label{fig:fig7}
\end{figure}

By plugging Equation~(\ref{eq:RCCO}) and Equation~(\ref{eq:SYZCO}) into Equation~(\ref{eq:throughput}) separately, we can estimate the total throughputs for both designs and further calculate  $OITs$ of all telescopes with the enhanced silver coating adopted,  (as shown in Figure~\ref{fig:fig7}(a) and~\ref{fig:fig7}(b)). As seen from this figure, we conclude that the SYZ telescope potentially has a less scientific productivity compared to the two-mirror telescope designed for any given FOV. Based on the $OIT$ curve for the RC design as shown in Figure~\ref{fig:fig7}(b), $OIT$ is increasing with FOV. However, for the SYZ system, although the $ENA$ of SYZ is smaller than RC system over a large FOV, the SYZ's $OIT$ will decrease with the increase of FOV after 30 arcmin, because of a very significant loss of throughput due to M4's central hole. 

\section{CONCLUSION}

In this paper, we systematically compare the scientific performance of telescopes with the two-mirror (RC) design to the SYZ design. % from the imaging quality and throughput over a large FOV. 
Our simulations show that RC telescopes have better diffraction imaging quality in the central FOV, and higher photon throughput due to less reflection loss and less obstruction loss compared to the SYZ telescopes. SYZ telescope is better than RC telescope in the diffraction limited imaging quality over a large FOV (20 arcmin). We further compare both telescope designs based on performance metrics of effective aperture and Central Intensity Ratio $CIR$. We find that SYZ design does not work as well as RC design under seeing limited observations, GLAO corrected observations and ExAO corrected observations. Furthermore, we propose a new performance evaluation criterion, i.e. the Observing Information Throughput ($OIT$) to evaluate the scientific productivity of telescopes. Using this OIT metric, we compare the two-mirror design with the SYZ design. The RC telescope has a better scientific performance compared to the innovative SYZ design. %In addition, we should also be cautious that SYZ system may have more cost and longer construction time due to its more complex structure, which we will discuss in our following paper. 

% \subsection{Figures and tables}

% Figures and tables should be placed at logical positions in the text. Don't
% worry about the exact layout, which will be handled by the publishers.

% Figures are referred to as e.g. Fig.~\ref{fig:example_figure}, and tables as
% e.g. Table~\ref{tab:example_table}.

% % Example figure

\section*{Acknowledgements}
We acknowledge valuable discussions from Jerry Nelson and Sandra Faber. We also acknowledge the great helps, supports and encouragements from Jiansheng Chen, Suijian Xue, Luis Ho, Lei Hao, Lu Feng. Finally, the authors dedicate this work to the memory of Jerry Nelson, without whom this work would not be possible.
%Support for this work was partially provided by NASA through Hubble Fellowship grant HST-HF2-51370 awarded by the Space Telescope Science Institute, which is operated by the Association of Universities for Research in Astronomy, Inc., for NASA, under contract NAS 5-26555.

%%%%%%%%%%%%%%%%%%%%%%%%%%%%%%%%%%%%%%%%%%%%%%%%%%



%\bibliographystyle{}


%%%%%%%%%%%%%%%%%%%%%%%%%%%%%%%%%%%%%%%%%%%%%%%%%%

%%%%%%%%%%%%%%%%% APPENDICES %%%%%%%%%%%%%%%%%%%%%

\appendix

\section{Central obstruction (CO) diffraction}
\label{sec:COdiff}

\begin{figure}
	\includegraphics[width=\columnwidth]{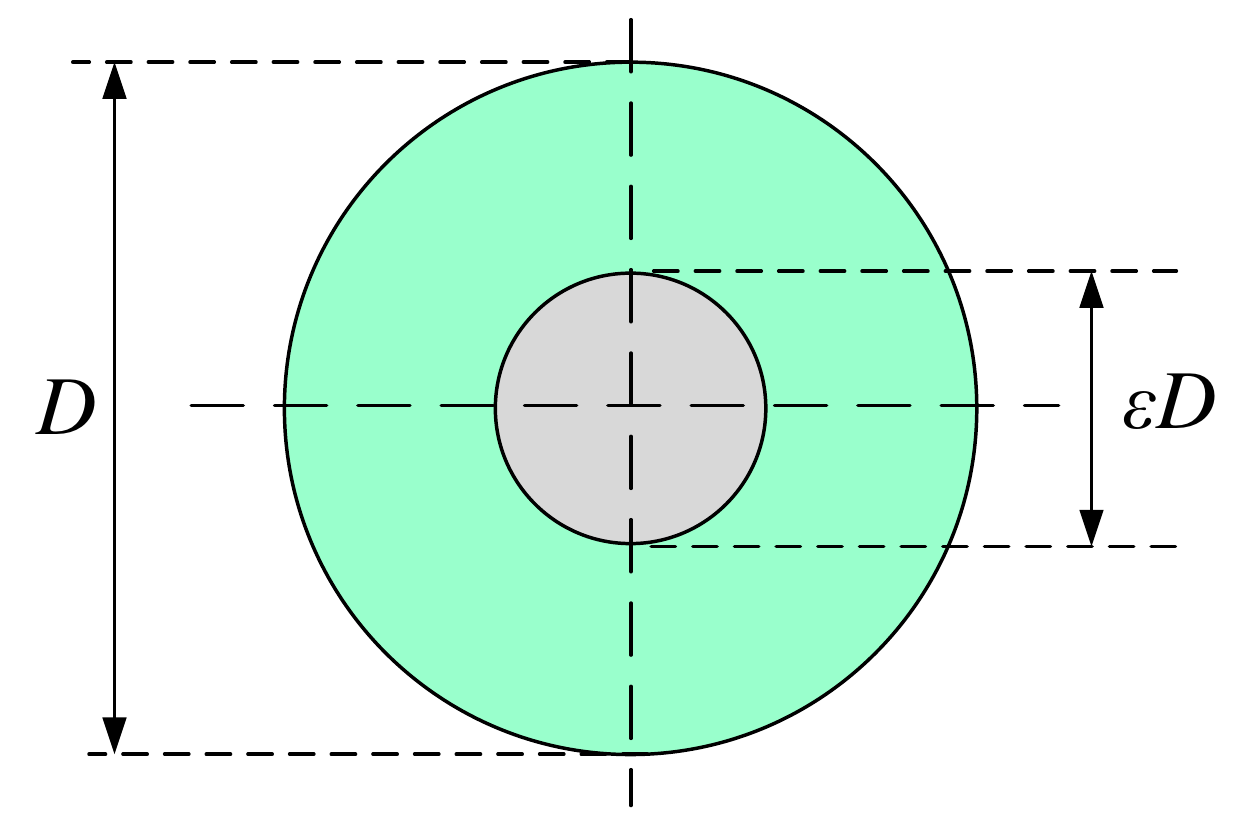}
    \caption{The central obstruction (CO): $D$=aperture diameter; $\epsilon$=relative size of CO in units of $D$.}
    \label{fig:figA1}
\end{figure}

In the SYZ telescope design, there are several central obstruction apertures along the optical path of SYZ telescope as show in Figure~\ref{fig:figA1}, which include, central obstruction caused by secondary mirror, central hole of primary mirror and central hole of fold mirror M4. In this appendix, we try to explore the diffraction patterns of the annular pupil with various central obscuration ratios, and analyze their effect on final image quality of optical systems.

Any obstruction placed in the light path of an imaging system will block a portion of the wavefront reaching the final focal surface. The consequence is a change in wave contribution at every point of the diffraction pattern, leading to a so-called obscured Airy pattern. The new intensity distribution of the obscured Airy patter can be described by (\citealt{Airy 1835}; \citealt{Sacek 2006}):

\begin{equation}
    I\left ( \theta  \right )=\frac{I_{0}}{\left ( 1-\epsilon ^{2} \right )^{2}}\left ( \frac{2J_{1}\left ( x \right )}{x}-\frac{2\epsilon J_{1}\left ( \epsilon x \right )}{x} \right )^{2},
	\label{eq:Intensity}
\end{equation}

\noindent
where $\epsilon$ is the annular aperture linear obscuration ratio, $I _{0}$ is the maximum intensity of the pattern at the Airy disc center with no obscuration, $J _{1}$ is the Bessel function of the first kind of order 1, and $x$ is defined as:

\begin{equation}
    x=ka\sin \theta \approx \frac{\pi R}{\lambda f/\#},
	\label{eq:xvar}
\end{equation}

\noindent
where $R$ is radial distance in the focal plane from the optical axis, $\lambda$ is the wavelength, $f/\#$ is the f-number of the system. In Equation~(\ref{eq:Intensity}), the central intensity with a given central obscuration (CO) ratio of $\epsilon$ is still normalized to $I _{0}$ by dividing a normalization factor of (1-$\epsilon ^{2}$)$^{2}$. As a result, the Strehl ratio (SR) of an optical system's PSF with CO ratio of $\epsilon$ can be written as:

\begin{equation}
    SR=(1-\epsilon ^{2})^{2}.
	\label{eq:SR}
\end{equation}

\noindent
Then by integrating Equation~(\ref{eq:Intensity}) without the normalization coefficient over image spot radius, the fraction of encircled energy is provided by:

\begin{eqnarray}
    E(R)&=& \frac{1}{1-\epsilon ^{2}} \{
1-J_{0}^{2}(x)-J_{1}^{2}(x)+\epsilon ^{2}\left [1-J_{0}^{2}(\epsilon x)-J_{1}^{2}(\epsilon x)  \right ] \nonumber  \\
&- &4\epsilon \int_{0}^{x}\frac{J_{1}(t)J_{1}(\epsilon t)}{t}dt\}.
	\label{eq:ER}
\end{eqnarray}

\begin{figure}
	\includegraphics[width=\columnwidth]{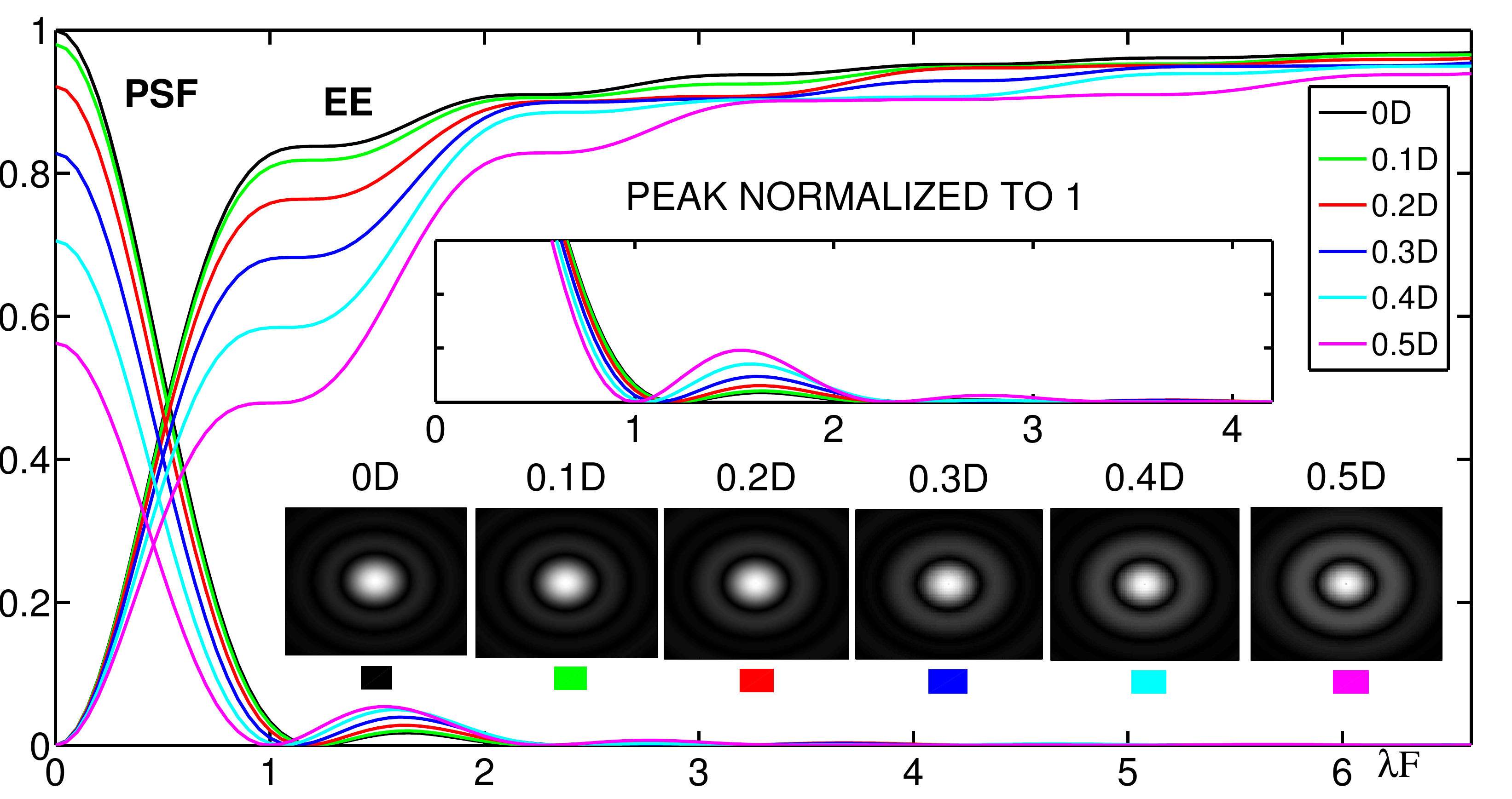}
    \caption{The effect of central obscuration on point-spread function (PSF) intensity distribution of optical systems.}
    \label{fig:figA2}
\end{figure}

Therefore, based on Equation~(\ref{eq:ER}), we can get the diffraction pattern of central obstruction and the corresponding point spread function and encircled energy distribution curve as shown in Figure~\ref{fig:figA2}. Due to the obstruction effect, the 80\% encircled energy (EE80) diameter of the spot with relative central obstruction of 0.5 is almost twice of the Airy disk size.

\section{Derivation of equivalent noise area (ENA)}
\label{sec:ENA}

Our derivation of $ENA$ largely follows the procedures in \citet{King 1983}, \cite{Mighell 2005}, and \citet{Mighell 2003}. Consider a CCD photometry of point sources. Assuming that we know the PSF of the observations, then a simple model of the observation can be written using the following parameters: stellar intensities ($E$) in photons, the coordinate of the point-source positions ($x$, $y$) in pixels, and observed background sky level in photons ($B$). If there are not only one star, and further, if some stars are overlapping, then the parameters of each star are dependent variables. The reasonable model of multiple overlapping PSFs will be non-linear. With the non-linear least-square fitting algorithm, the non-linear function can simultaneously determine any dependent or independent parameters in the non-linear model functions (\citealt{Mighell 2003}). 

Following \cite{Mighell 2005}, assuming that we have a CCD images with $N$ pixels and that $n _{i}$ is the number of photons recorded in the $i _{th}$ pixel. The $i _{th}$ pixel resides in the position of ($x _{i}$, $y _{i}$) of this CCD, and this pixel has an error of $\sigma _{i}$ photons. Further, let us denote $m$($x$, $y$; $n _{1}$,..., $n _{M}$) to be an observational model of the pixel values in CCD that has the coordinate ($x$, $y$). Let us denote vector \textbf{p} to represent all the model parameters [\textbf{p} $\equiv$ ($p _{1}$,..., $p _{M}$)]. The observational model of $i _{th}$ pixel can be written as $m _{i}\equiv m$($x$, $y$; \textbf{p}).

We use $\chi ^{2}$ to quantitatively measure the goodness of the fit between a non-linear model and data, in which $\chi ^{2}$ can be expressed as follows: 

\begin{equation}
    \chi ^{2}(p)=\sum_{i=1}^{N}\frac{1}{\sigma _{i}^{2}}\left ( z_{i}-m_{i} \right )^{2}.
	\label{eq:Chi}
\end{equation}

\noindent
Assuming \textbf{p}$_{0}$ is the optimal parameter vector, let us consider a parameter \textbf{p}, and then the standard errors according to the non-linear least-square fitting is:

\begin{equation}
    \delta _{j}=\left ( \sum_{i=1}^{N}\frac{1}{\sigma _{i}^{2}}\frac{\partial m_{i}}{\partial p_{j}} \right )^{-1/2}.
    \label{eq:deltaj}
\end{equation}

\noindent
The observational model for $i _{th}$ pixel would be

\begin{equation}
    N_{i}=B+E\phi _{i},
	\label{eq:Npixel}
\end{equation}

\noindent
where $E$ is the total number of photons received from the source, and $\phi _{i}$ is the value of $i _{th}$ pixel of the normalized PSF. For faint limit, the error of $E$ can be expressed as follows based on Equation~(\ref{eq:deltaj}):

\begin{equation}
    \sigma _{E}^{2}=\left ( \sum_{i=1}^{N}\frac{1}{\sigma _{rms}^{2}\left ( \frac{\partial }{\partial E} E\phi \right )^{2}} \right )^{-1}.
	\label{eq:SigmaOne}
\end{equation}

\noindent
Consider the large array of CCD and the discrete CCD array can be approximated as continuous, then Equation~(\ref{eq:deltaj}) can be written as follows: 

\begin{equation}
    \sigma _{E}^{2}=\sigma _{rms}^{2}\left [ \iint_{ }^{ } \phi ^{2}dxdy\right ]^{-1}.
	\label{eq:SigmaTwo}
\end{equation}

\noindent
Thus, in order to get a complete information of the photon from the source ($E$), we need to figure out $[ \iint_{ }^{ } \phi ^{2}dxdy]^{-1}$. Thus, the $[ \iint_{ }^{ } \phi ^{2}dxdy]^{-1}$ is defined as Equivalent-Noise-Area ($ENA$).

%%%%%%%%%%%%%%%%%%%%%%%%%%%%%%%%%%%%%%%%%%%%%%%%%%

% Don't change these lines
\bsp	% typesetting comment
\label{lastpage}
\end{document}